# Unraveling phase transformation with phononic hyperbolicity using off-resonant terahertz light

Hanli Cui, Jian Zhou[*]

*Center for Alloy Innovation and Design, State Key Laboratory of Mechanical Behaviors of Materials, Xi'an Jiaotong University, China 710049*

[*]Email: jianzhou@xjtu.edu.cn

Abstract

Noncontacting and nondestructive control of geometric phase in conventional semiconductors plays a pivotal role in various applications. In the current work, we present a theoretical and computational investigation on terahertz (THz) light-induced phase transformation of conventional binary semiconducting compounds among different structures including rock-salt, zinc-blende, wurtzite, and hexagonal phases. Using MgS and MgSe as prototypical examples, we perform anharmonic phonon mediated calculations and reveal large contrasting lattice contributed dielectric susceptibility in the THz regime. We then construct a THz-induced phase diagram under intermediate temperature and reveal rock-salt to hexagonal and then wurtzite structure transformations with increasing light intensity. This does not require a high temperature environment as observed in traditional experiments. The low energy barrier suggests that the phase transition kinetics can be fast, and the stable room temperature phonon dispersions guarantee their non-volatile nature. Furthermore, we disclose the phononic hyperbolicity with strong anisotropic THz susceptibility components, which serves as a natural hyperbolic material with negative refractive index. Our work suggests the potential to realize metastable hidden phases using noninvasive THz irradiation, which expands the conventional pressure-temperature ($P - T$) phase diagram by adding light as an additional control factor.

## I. Introduction

Optomechanical responses and behaviors have been receiving hectic attention over the past few years, owing to their ultrafast kinetics, precise control nature, and non-invasive character [1]. The contactless feature would limit the introduction of foreign atom impurities, and reduce the complicated fabrication of electrochemical electrodes on the samples. Owing to these advances, light-induced structural deformation and phase transformation are considered to be useful for various applications, such as opto-actuators, security check, food quality scan, and next generation data-storage techniques [2-6]. Within the electromagnetic wave spectrum, the terahertz regime (within about 0.1–10 THz) has been less investigated and used, leaving a gap between the shorter wavelength (infrared and visible light) and longer wavelength (micro and radio wave) regimes [7]. Recent years have witnessed a burst of exploration and scrutinization of THz light-matter interactions, due to the vast development of its sources. THz optics have been considered to be athermic to materials, providing a good platform for non-destructive and reversible structural control [8-11]. For example, two independent experiments have shown that THz light could trigger phase transformation of $SrTiO_3$ from its quantum paraelectric phase to a hidden ferroelectric structure [12, 13]. These demonstrate that in addition to the conventional phase control with pressure ($P$) and temperature ($T$), light has been another tuning dimension, which is less susceptible to lattice damage and can be flexibly modulated.

In spite of these efforts, the THz induced optomechanical responses are still in their early stage. Since the THz light could resonantly excite infrared (IR)-active vibration modes, while the phase transformation is usually subject to Raman-active modes, many prior studies focus on anharmonic interactions between them, such as unravelling hidden phase in perovskite and inducing Weyl semimetals in Dirac semimetals [14-16]. However, the resonant IR-active mode excitations correspond to light absorption, and the potential over-heating problem naturally exists. This further requires that only a few THz pulses can be applied to the samples, rather than continuous wave irradiation. Off-resonant light-induced phase transformations could avoid such a challenge, which corresponds to a scattering process, during which only the Raman-active phonons are excited and the IR-active modes remain silent [17].

In the current work, we perform first-principles calculations to show that an intermediate THz light centered at ~1 THz (below all IR-active modes) could trigger structural phase transformations of conventional bulk binary semiconductors. Taking a prototypical Mg$X$ ($X$ = S and Se) as examples, we suggest that THz light could control its phase from the ground state rock-salt structure to hexagonal and then wurtzite phases, depending on the temperature and light intensity [18-20]. The optical responses are carefully evaluated from both electron and lattice vibration contributions, with finite temperature effect included according to anharmonic vibrational interactions up to the fourth order. We show that the off-resonant THz light could serve as an alternative route to phase control, which might replace the conventional thermal and mechanical approaches with faster kinetics. Furthermore, we propose that the anisotropic wurtzite and hexagonal phases host robust natural hyperbolicity in the THz frequency regime. Hyperbolic dispersion arises in systems once their dielectric function components exhibit opposite signs different principal axes. They possess various optical benefits such as low-loss properties, broadband electromagnetic wave response, all-angle negative refraction, deep subwavelength imaging, and outstanding light confinement capabilities. Artificially engineered hyperbolic metamaterials usually require complicated fabrication and precise control of their pattern. On the contrary, natural hyperbolic materials could exhibit these characteristics at specific frequency ranges in a single material, so that they have been attracting tremendous attention in recent years [21-24]. Until now, various natural hyperbolic materials have been theoretically and experimentally proposed, such as graphite [25], transition metal dichalcogenide [26], α-$MoO_3$ [27], and black phosphorus [28]. Topological materials have also been predicted as natural hyperbolic materials with a wide hyperbolic frequency range that spans across the near-infrared to visible spectrum [29]. In these cases, the hyperbolic frequencies are mainly focused on the visible to ultra-violet regimes [30-33]. Here, we show that non-layered structure could also host hyperbolic features with negative light refractive index, which arises from anisotropic phonon polaritons in the THz regime. The development of THz hyperbolic materials could facilitate various cutting-edge technology such as good control of the THz wave, precise imaging, and non-destructive quality check.

## II. Computational Methods

Our first-principles calculations are performed using the density functional theory (DFT) with its exchange correlation interaction treated in the form of the generalized gradient approximation (GGA) method in the solid state Perdew-Burke-Ernzerhof (PBEsol) form [34], as implemented in the Vienna *ab initio* Simulation Package (VASP) code [35]. The projector augmented-wave (PAW) method is adopted to treat the core electrons [36, 37], while the valence electrons are represented by plane wave basis set with a kinetic cutoff energy of 500 eV. The total energy and force convergence criteria are set to be $1\times10^{-8}$ eV and $1\times10^{-7}$ eV/Å, respectively. The first Brillouin zone (BZ) is represented by Γ-centered Monkhorst-Pack ***k***-mesh of (9×9×9) grids [38]. In order to incorporate finite temperature effect of phonon dispersion and phonon vibration lifetime, we apply the self-consistent approach (SCPH scheme [39]) with microscopic anharmonic force constants extracted from density-functional calculations using the least absolute shrinkage and selection operator technique [40]. To generate random samples and compute force constants, *ab initio* molecular dynamics (AIMD) simulations are performed within the (2×2×2) supercells [41]. Non-analytic correction by the Ewald method and the Born effective charge tensors evaluation are adopted in the framework of the density-functional perturbation theory (DFPT).

### III. Results and Discussions

#### A. Geometric structure of MgS and MgSe

We illustrate the off-resonant THz induced phase transformation using simple binary compound semiconductors MgS and MgSe, which have been experimentally examined over five decades [42]. Each of them possesses a ground state at the room temperature in the rock-salt phase, belonging to the $Fm\bar{3}m$ space group [18, 43]. In order to examine their relative phase stability, we investigate other potential structures including the wurtzite (space group $P6_3mc$), zinc-blende (space group $F\bar{4}3m$), and hexagonal close packed (space group $P6_3/mmc$) structures, as plotted in Fig. 1. After relaxation at zero temperature, we list their lattice parameters and relative energies in Table I. One sees that the rock-salt $Fm\bar{3}m$ takes the lowest energy among all these phases, consistent with experimental facts. In addition, as the Mg–$X$ ($X$ = S and Se) bonds are strongly ionic, these phases show semiconducting electronic band structures with large band gap values (> 1.75 eV, see Fig. S1 in Supplemental Material, SM [44]). Hence, we adopt the independent particle approximation to estimate their electron-contributed dielectric function [45],

$$\varepsilon_{ij}^{el}(\omega) = \delta_{ij} - \frac{e^2}{\varepsilon_0} \int_{BZ} \frac{d^3\mathbf{k}}{(2\pi)^3} \sum_{c,v} \frac{\langle u_{v\mathbf{k}}|\nabla_{k_i}|u_{c\mathbf{k}}\rangle \langle u_{c\mathbf{k}}|\nabla_{k_j}|u_{v\mathbf{k}}\rangle}{\hbar\left(\omega_{c\mathbf{k}}-\omega_{v\mathbf{k}}-\omega-\frac{i}{\tau^{el}}\right)} \qquad (1)$$

Here, $\varepsilon_0$ is the vacuum permittivity and $\delta_{ij}$ represents Kronecker delta symbol. $|u_{v\mathbf{k}}\rangle$ and $\omega_{v\mathbf{k}}$ represent periodic part of the Bloch wavefunction and its eigenfrequency in the valence (or conduction) band at momentum $\mathbf{k}$. At the THz regime, the dielectric function value is insensitive to the electron lifetime $\tau^{el}$ (see Fig. S2 [44] for details), which are also tabulated in Table I.

Table I. Space group, optimized lattice constant ($a$, $b$, $c$), bond length ($d_{\text{Mg-X}}$), relative energies for MgS and MgSe at zero temperature, and electron contributed dielectric function components at the THz regime ($\varepsilon_{xx}^{el} = \varepsilon_{yy}^{el}$) of MgS and MgSe compounds.

| Materials | Space group | Lattice constant (Å) | $d_{\text{Mg-X}}$ (Å) | Relative energy (eV/f.u.) | $\varepsilon_{xx}^{el} = \varepsilon_{yy}^{el}$, $\varepsilon_{zz}^{el}$ |
|---|---|---|---|---|---|
| MgS | $Fm\bar{3}m$ | $a=b=c=5.22$ | 2.61 | 0.00 | 5.82 |
| | $P6_3mc$ | $a=b=4.05$ $c/a=1.60$ | 2.47 | 0.06 | 4.32, 4.41 |
| | $F\bar{4}3m$ | $a=b=c=5.69$ | 2.46 | 0.03 | 4.36 |
| | $P6_3/mmc$ | $a=b=3.66$ $c/a=1.69$ | 2.62 | 0.11 | 5.81, 5.91 |
| MgSe | $Fm\bar{3}m$ | $a=b=c=5.50$ | 2.75 | 0.00 | 7.04 |
| | $P6_3mc$ | $a=b=4.26$ $c/a=1.61$ | 2.60 | 0.08 | 4.95, 5.04 |
| | $F\bar{4}3m$ | $a=b=c=5.99$ | 2.59 | 0.03 | 4.99 |
| | $P6_3/mmc$ | $a=b=3.87$ $c/a=1.67$ | 2.76 | 0.13 | 7.07, 7.14 |

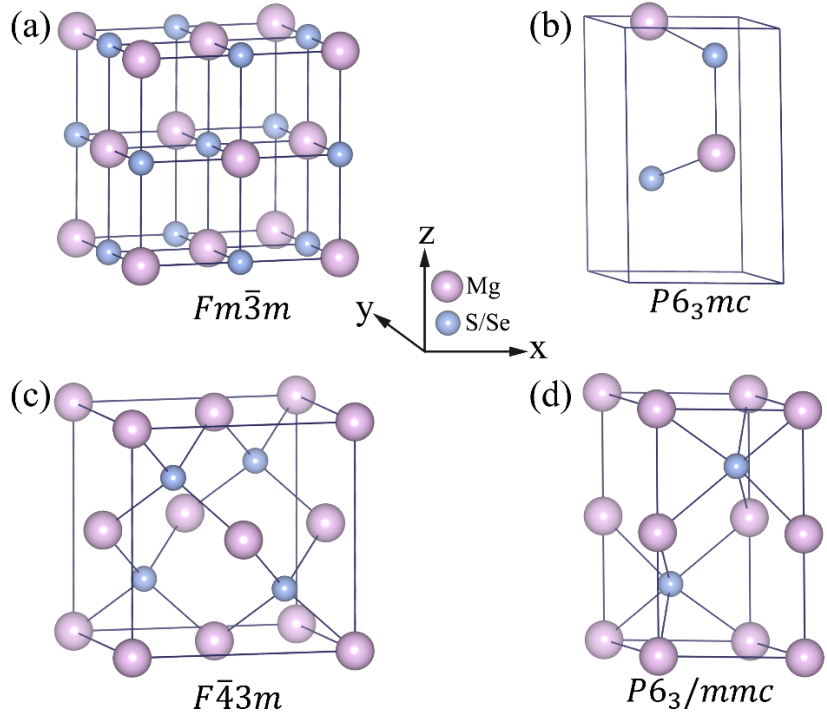

FIG. 1. Atomic structures of MgS and MgSe in the (a) $Fm\bar{3}m$ (rock-salt), (b) $P6_3mc$ (wurtzite), (c) $F\bar{4}3m$ (zinc-blende), and (d) $P6_3/mmc$ (hexagonal closed packed) phases.

### B. Frequency dependent dielectric function at THz regime

Next, we calculate their phonon dispersions under finite temperature, as plotted in Fig. 2. As the compounds are ionic, we include the longitudinal optical (LO) and transverse optical (TO) splitting effect [46, 47]. The dashed curves represent the conventional harmonic approximation results, and the colored solid curves are obtained from anharmonic corrections. One sees that the largest frequencies for MgS phases are ~12 THz (see Fig. S3 [44] for MgSe phonon dispersions, giving largest frequencies of ~11 THz). For the $P6_3mc$ and $P6_3/mmc$ phases, significant LO-TO splitting near the $\Gamma$ point occurs, while for it is much smaller for the high symmetric $Fm\bar{3}m$ and $F\bar{4}3m$. In both materials, the thermal effect arising from anharmonic vibrations generally enhances the optical branch frequency in the $Fm\bar{3}m$ and $P6_3mc$ phases, while it shows slight corrections for $F\bar{4}3m$ and $P6_3/mmc$. In each panel, we mark the dominant IR-active modes that are subject to the ion contributed dielectric functions.

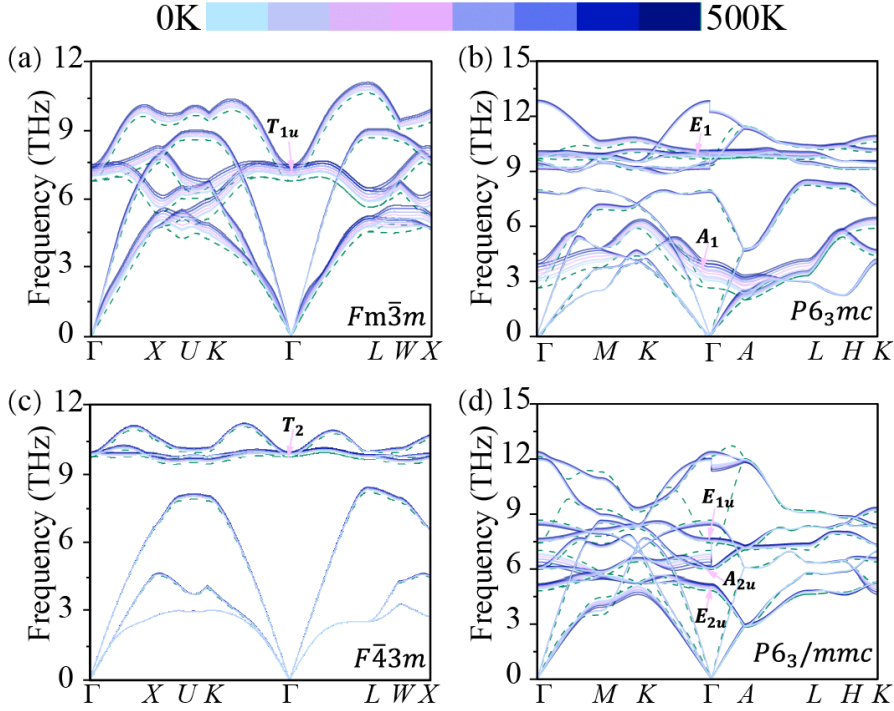

FIG. 2. Phonon dispersion relation under harmonic vibration approximation (dashed curves) and corrected by anharmonic interactions (solid curves) under finite temperature for MgS. (a)–(d) are $Fm\bar{3}m$, $P6_3mc$, $F\bar{4}3m$, and $P6_3/mmc$, respectively. We mark the dominant IR-active modes for each panel, which contribute to the dielectric function spectrum.

With the phonon dispersion relation under finite temperature, we compute the ionic vibration contributed dielectric function according to linear response theory [46]

$$\varepsilon_{ij}^{\text{ion}}(\omega) = \frac{1}{\Omega}\sum_m \frac{S^*_{m,ij}}{\omega_m^2 - \left(\omega + \frac{i}{\tau_m^{\text{ion}}}\right)^2}, \qquad (2)$$

where $\Omega$ is the total volume of the unit cell, $\omega_m$ and $\tau_m^{\text{ion}}$ are the eigenfrequency and phonon lifetime for each mode $m$ near the $\Gamma$ point, respectively. $S^*_{m,ij} = \sum_{\kappa\kappa',i'j'} z^*_{\kappa,ii'} z^*_{\kappa',jj'} u_{m\kappa,i'} u_{m\kappa',j'}$ is the mode-oscillator strength tensor, with $z^*_{\kappa,ii'}$ and $u_{m\kappa,i'}$ are the Born effective charge component and displacement mode for ion-$\kappa$ [48]. The Born effective charge components are calculated according to harmonic approximations, as the electron-phonon coupling is not significantly large. Their results are listed in Table S1 [44]. The lifetime $\tau_m^{\text{ion}}$ for each mode can be estimated using anharmonic interactions [39]. As shown in Figs. 3(a) and 3(c), we plot the lifetime variation of each IR-active mode for MgS and MgSe under different temperature. Their corresponding eigenfrequency variations are plotted in Figs. 3(b) and 3(d), which slightly enhances with temperature. In this case, we assume a perfect crystal structure without impurities, and the electron-phonon coupling effects are not included. We find that the IR-active mode could last over a few to a few tens of picoseconds until being scattered. As temperature increases, the lifetime values reduce, consistent with the stronger scattering process under heating effect. For example, one sees at low temperature (100 K), the $A_1$ mode in $P6_3mc$ phase shows a significantly long lifetime (over 40 ps). This would result in a sharp peak for imaginary part of the dielectric function (see below). It quickly reduces to 11 and 6 ps at room temperature (300 K) and higher temperature of 500 K, respectively. This strong temperature dependence is also consistent with the large temperature renormalization for its phonon dispersion [Fig. 2(a)]. When the IR-active mode lies near other modes in the frequency regime, it might experience stronger anharmonic interactions. For example, the $E_{1u}$ mode in $P6_3/mmc$ phase possesses a much shorter lifetime (1 ps at room temperature). This would yield a broader peak in the imaginary part of dielectric function.

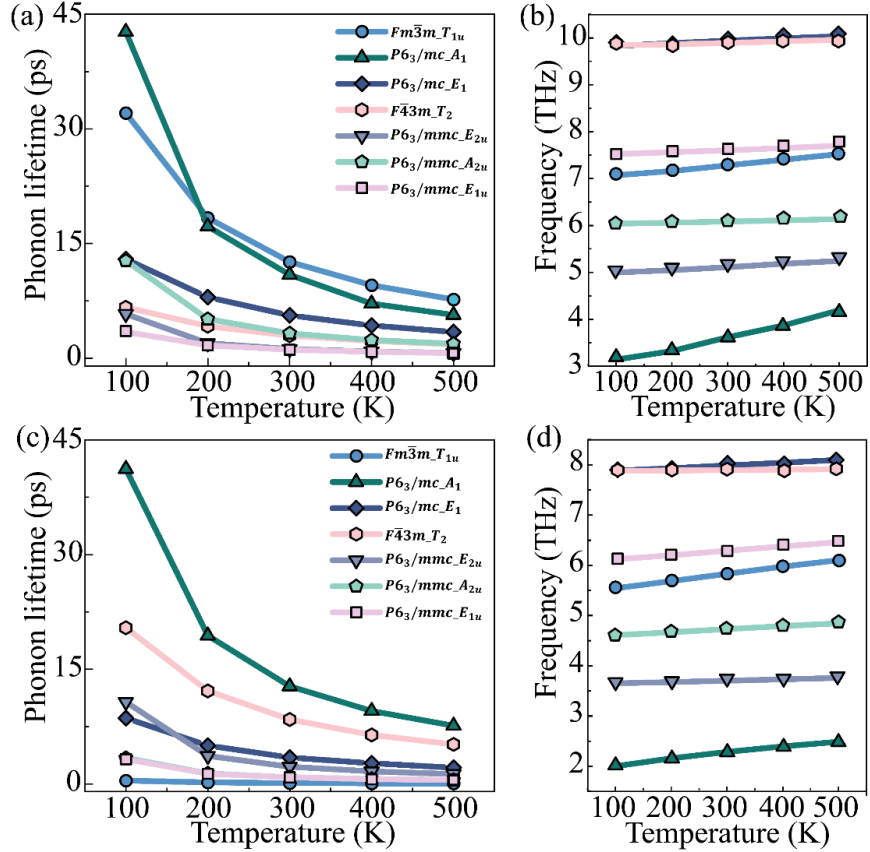

FIG. 3. (a) Lifetime and (b) eigenfrequency variations under finite temperature of the IR-active modes of MgS. (c) and (d) are the corresponding results for MgSe.

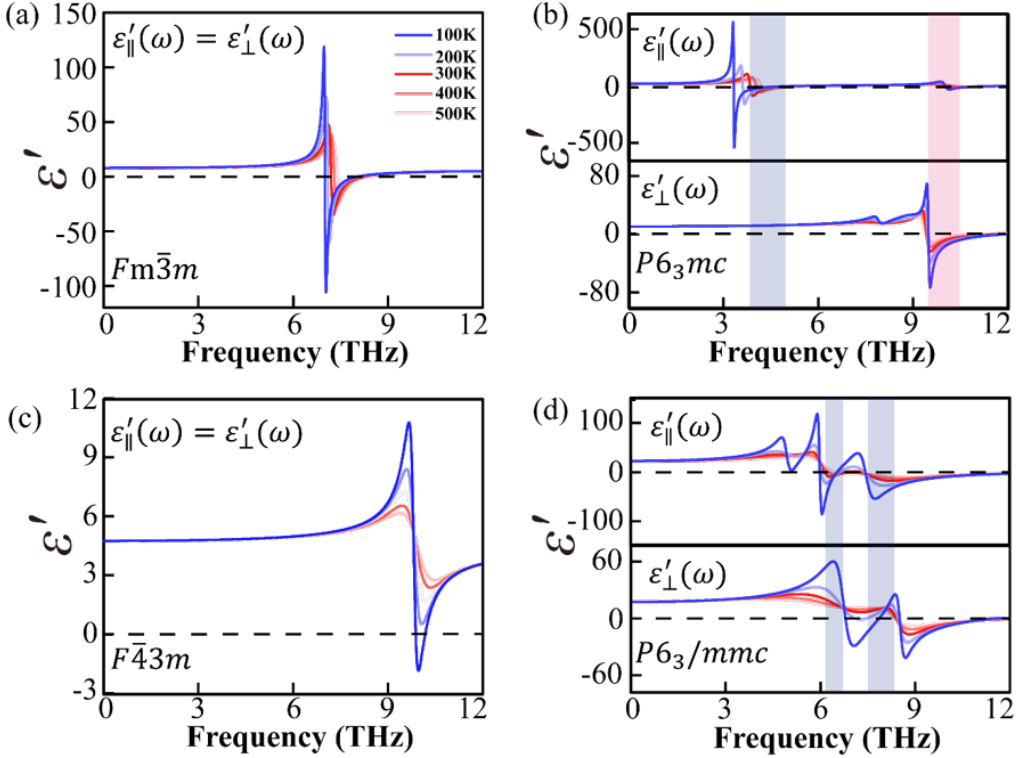

FIG. 4. Real part of total dielectric function in the THz regime for MgS in (a) $Fm\bar{3}m$, (b) $P6_3mc$, (c) $F\bar{4}3m$, and (d) $P6_3/mmc$ phases. Here, temperature effect is taking into account for their modulations on the eigenfrequency and phonon lifetime of IR-active modes. Shaded in (b) and (d) represent phononic hyperbolicity regimes (light red for type-I and light blue for type-II) at room temperature.

We plot the real part of total dielectric function $\varepsilon_{ij}^{tot,\prime}(\omega) = \varepsilon_{ij}^{el,\prime}(\omega) + \varepsilon_{ij}^{ion,\prime}(\omega)$ for each phase of MgS in Fig. 4, and their corresponding imaginary parts are shown in Fig. S4 [44]. Due to the presence of $C_{4z}$ rotation in all phases, $\varepsilon_{xx} = \varepsilon_{yy}$, which will be denoted as $\varepsilon_\parallel$. The $\varepsilon_{zz}$ is then denoted as $\varepsilon_\perp$. From Eq. (2), only IR-active modes give a peak in the imaginary part. Long lifetime corresponds to a sharp and narrow peak, while a reduced lifetime broadens them. According to the Kramers-Kronig relation, the real part of dielectric function shows several jumpy fluctuations, at the peak frequencies of its corresponding imaginary part. The smearing in the real part of dielectric function is consistent with the imaginary part, stemming from the lifetime $\tau_m^{ion}$. When the frequency is well-below the IR-active mode frequency, the imaginary part is almost zero, indicating no direct light absorption. The real part of dielectric function, on the other hand, is nonzero and keeps a constant. Note that this value is almost insensitive to temperature (IR-active frequency and lifetime). One sees that at low frequency, the $F\bar{4}3m$ has the smallest dielectric function value, $\varepsilon_\parallel^{tot,\prime} = \varepsilon_\perp^{tot,\prime} = 5.74$ at frequency of 1 THz. This is due to the dual effects of its high IR-active frequency ($\omega_{IR} = 9.94$ THz) and small Born effective charge value ($z^*_{Mg,xx} = z^*_{Mg,zz} = 1.90$). In general, the dielectric function is enhanced with low $\omega_{IR}$ and large $z^*$. For example, the $P6_3mmc$ possesses $\varepsilon_\parallel^{tot,\prime} = 20.90$ and $\varepsilon_\perp^{tot,\prime} = 18.30$ at 1 THz, while they furthermore reach $\varepsilon_\parallel^{tot,\prime} = 25.40$ and $\varepsilon_\perp^{tot,\prime} = 19.52$ in the $P6_3mc$ phase. Recently, similar wurtzite BeO is suggested to possess large ion contributed dielectric function value at low frequency limit [49], in which the physical picture is fully consistent with our current results.

### C. Terahertz light induced phase transformation

According to our previous works [50], the real part of dielectric function variation in different phases (alternatively, the gradient of the reaction coordinate system) serves as a generalized driving force for nonresonant light-induced phase transformation. This is akin to the optical tweezer technique and is an athermic process. Microscopically, this belongs to the impulsive stimulated Raman scattering and our previous prediction [50] on THz-induced MoTe2 phase transition are well-consistent with subsequent experimental observations [51]. Without direct light absorption, the Gibbs free energy (GFE) density (per f.u.) would vary according to

$$\Delta \mathcal{G}(E) = -\frac{1}{4}\varepsilon_0 \varepsilon_{ii}^{tot,\prime}(\omega) E_i E_i^* \qquad (3)$$

Here, the light alternating electric field is denoted as $\mathcal{E}(\omega, t) = \mathrm{Re}(\mathbf{E}e^{i\omega t})$ [15]. In the current work, we focus on using $z$-polarized THz light, as the $\varepsilon_{zz}^{tot,\prime}(\omega)$ varies largely among different phases. Then we track the GFE density variation as a function of both temperature and light field.

We plot THz-dressed phase diagram for MgS and MgSe, taking the incident light frequency of 1 THz. This is a typical value, and can be slightly adjusted as long as it does not enhance up to the IR-active mode frequency. Also note that it cannot be reduced to very low frequency, as additional sources of dielectric medium (such as orientation and dipolar effect, usually $\sim 10^5 - 10^{10}$ Hz) may take part into the effect, which are not considered here. The thermal effect is taking into account via evaluating the free energy of electron and ion subsystems,

$$F_{el}(T) = -2k_B T \sum_{n\mathbf{k}}[f_{n\mathbf{k}} \ln f_{n\mathbf{k}} + (1 - f_{n\mathbf{k}}) \ln(1 - f_{n\mathbf{k}})] \qquad (4)$$

and

$$F_{\text{ion}}(T) = \frac{1}{2}\sum_{mq} \hbar\omega_{mq} + k_B T \sum_{mq} \ln\left[1 - e^{-\frac{\hbar\omega_{mq}}{k_B T}}\right] \quad (5)$$

Here, $f_{nk}$ is the Fermi-Dirac distribution for electronic band index $n$ at $\mathbf{k}$. The total GFE is then

$$\mathcal{G}(E,T) = U + F_{\text{el}}(T) + F_{\text{el}}(T) + \Delta\mathcal{G}(E). \quad (6)$$

Here, $U$ is DFT calculated internal energy of each phase. As seen in Fig. 5, when the THz electric field strength is below 0.7 V/nm (corresponds to light intensity of $6.13 \times 10^{10}$ W/cm$^2$), the $Fm\bar{3}m$ phase remains to be the ground state in both compounds. Above this critical value, the $P6_3/mmc$ and $P6_3mc$ become more stable. The $P6_3/mmc$ covers a larger region in MgS than in MgSe, where the $P6_3mc$ directly appears under low temperature. Note that such an intensity of THz field has been achieved in experiments using the free electron laser technique [52]. In this regard, we suggest that THz irradiation could be used to drive phase transformation in both MgS and MgSe compounds, from an isotropic structure to two anisotropic structures. The $F\bar{4}3m$, on the other hand, cannot be directly triggered via THz light illumination, owing to its small dielectric function value. We note that both $P6_3/mmc$ and $P6_3mc$ are metastable without imaginary vibration modes in the whole first BZ, such phase transformations are nonvolatile when THz light is removed. Hence, contrary to conventional resonant light induced phase transformation that light absorption is limited by the penetration depth, here the sample is almost transparent under 1 THz (with $\varepsilon'' \simeq 0$) so that the phase transformation would occur collectively in the whole material. In this way, the nucleation and growth mechanism could be eliminated, especially when the energy barrier is not significant under finite temperature. We use the climb image nudged-elastic-band method to estimate the barrier from $Fm\bar{3}m$ to $P6_3/mmc$ and $P6_3mc$, giving 0.32 eV/atom and 0.39 eV/atom for MgS (or 0.34 eV/atom and 0.31 eV/atom for MgSe). Note that under THz irradiation, these values may be furthermore reduced, but a rigorous computation is not straightforward as the definition of phonon dispersion on the transformation path is not well-defined. Note that the phase transformations in bulk materials are usually reversible, potentially owing to the residual back-stress in the system. In the current case, as the phase transformation is expected to occur collectively, such a problem could be greatly eliminated. In order to further reduce such an effect, one may consider using nanoparticles in practical experiments. In fact, recent experiments have demonstrated irreversible phase transformations in nanoscale transition metal dichalcogenide monolayers [51].

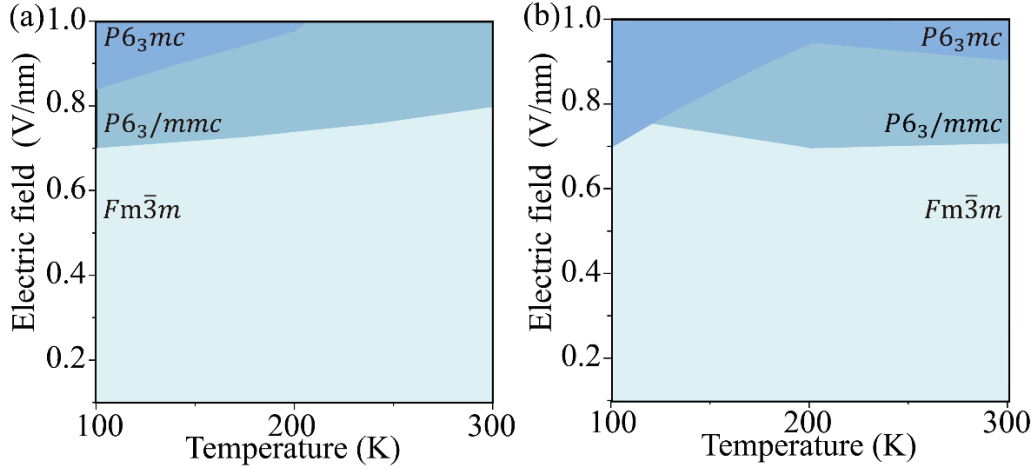

FIG. 5. Phase diagram under intermediate temperature and THz electric field strength (at the frequency of 1 THz) for (a) MgS and (b) MgSe. The alternating electric field is polarized along the z direction.

### *D. Phononic hyperbolicity and negative refraction in THz regime*

The strongly anisotropic dielectric function in THz-induced $P6_3/mmc$ and $P6_3mc$ phases could lead to various exotic optical responses. Here, we propose that they both exhibit hyperbolic dispersion relation, serving as phononic natural hyperbolic materials [Fig. 6(a)]. We note that previously studied electronic-based natural hyperbolic materials are usually metals, in which the intraband electronic transition (usually described by the well-known Drude model) is strongly susceptible to the carrier relaxation time ($\tau^{el}$). Therefore, the hyperbolic frequency regime strongly depends on its specific value. In principle, this relaxation time is a function of band index and electronic momentum, and its exact value is not straightforwardly determined. This poses large uncertainties of the predicted hyperbolic frequency regime, mainly depending on impurity levels and environmental conditions. In addition, the sizable $\varepsilon^{el,''}(\omega)$ at this regime leads to non-negligible energy loss. Such an energy loss would set a limit for maximum wave vector ($k_{max}$). On the contrary, in the THz regime, these limitations can be largely reduced, as the $\varepsilon^{el,''}(\omega) \simeq 0$ when photon frequency is well below the semiconductor bandgap. Furthermore, the band-resolved ionic vibration lifetime values $\tau_m^{ion}$ can be accurately computed according to anharmonic interactions.

According to previous works, the calculated complex dielectric function (and THz reflectivity) could exhibit good consistency with experimental measurements [53-55]. Our results on rock-salt MgS also agree well with previous work [56] which shows accurate reflectivity spectrum with experimental results. In this regard, our predicted hyperbolic dispersion frequency regime and negative refractive index (see below) could be directly verified by experiments.

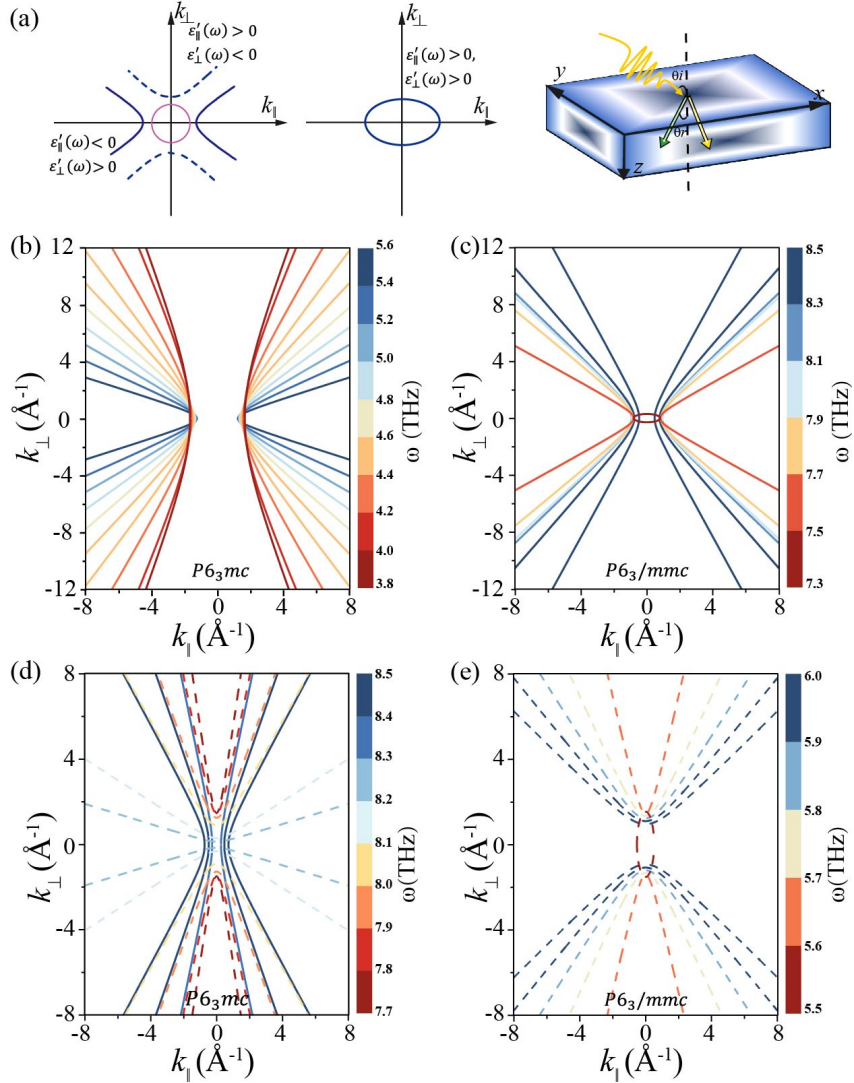

FIG. 6. (a) Schematic plots for phononic (left panel) hyperbolic dispersion (type-I, dashed curve, type-II, solid curve, pink curve, isotropic medium such as air) and (middle panel) elliptical dispersion. The right panel represents positive (yellow arrow) and negative (green arrow) refraction of the transverse wave. Isofrequency contour of MgS in (b) $P6_3mc$ and (c) $P6_3/mmc$ phases, and MgSe in (d) $P6_3mc$ and (e) $P6_3/mmc$, at room temperature.

The hyperbolic dispersion indicates that $\varepsilon'_\parallel(\omega) \times \varepsilon'_\perp(\omega) < 0$, according to $\frac{k_\parallel^2}{\varepsilon'_\parallel(\omega)} + \frac{k_\perp^2}{\varepsilon'_\perp(\omega)} = \frac{\omega^2}{c^2}$, where $c$ refers to the speed of light [57]. If $\varepsilon'_\perp(\omega) < 0$ and $\varepsilon'_\parallel(\omega) > 0$, the hyperbolic nature is categorized as type-I, and when $\varepsilon'_\perp(\omega) > 0$ and $\varepsilon'_\parallel(\omega) < 0$, it is type-II [58, 59]. We mark these frequency regimes of MgS in Figs. 4(b) and 4(d) at room temperature, and the MgSe case can be seen in Fig. S5 [44]. One sees that the hyperbolic frequency spans from 3.8 to 5.6 THz and 7.4 to 8.5 THz in MgS-$P6_3mc$. For example, at the frequency of 3.9 THz, we have $\varepsilon_\parallel(\omega) = -86.7 + 80.4i$ and $\varepsilon_\perp(\omega) = 9.7 + 0.5i$, indicating a type-II dispersion. At a higher frequency of 9.7 THz, the signs of their real parts are flipped, giving a type-I one with $\varepsilon_\parallel(\omega) = 17.9 + 14.2i$ and $\varepsilon_\perp(\omega) = -20.0 + 17.9i$. Similar results can be seen in the $P6_3/mmc$ phase. We plot the temperature dependent hyperbolic frequency window in Fig. 7. Under low temperature, the hyperbolic frequency window can be furthermore enlarged. In Figs. 6(b) and 6(c) we plot the isofrequency contours for MgS in $P6_3/mmc$ and $P6_3mc$ phases [with the case of MgSe shown in Figs. 6(d) and 6(e)]. If the THz light is incident from isotropic medium [such as air, corresponding to the pink circle in Fig. 6(a)] into the sample, any $k_\parallel$ would corresponds to a $k_\perp$ (almost infinite $k_{max}$ with marginal light absorption), indicating that the light can propagate in a long distance without a significant cutoff and suppression.

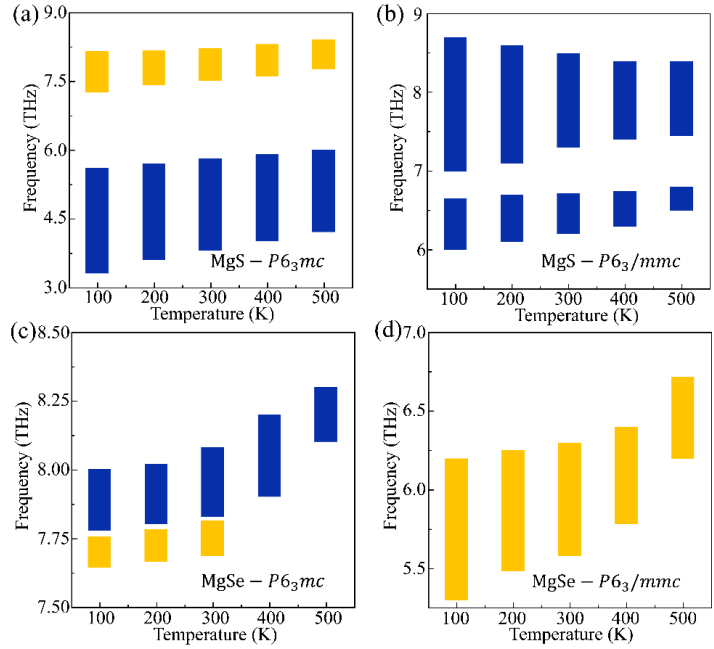

FIG. 7. Temperature-dependent hyperbolic frequency window (type-I, yellow bar, type-II, blue bar) of MgS in (a) $P6_3mc$ and (b) $P6_3/mmc$, and MgSe in (c) $P6_3mc$ and (d) $P6_3/mmc$ phases.

We also estimate the negative refraction nature at the interface between air and the (001) surface of MgS (or MgSe). Under an incident angle of $\theta_i$, the refraction angle is [60]

$$\theta_r = \arctan\left(\frac{S_\parallel}{S_\perp}\right) = \arctan\left(\frac{\sqrt{\varepsilon_\perp}\sin\theta_i}{\sqrt{\varepsilon_\parallel^2 - \varepsilon_\parallel \sin^2\theta_i}}\right) \quad (7)$$

Here, $S_i = \frac{\varepsilon_{ij}k_j}{2\omega\varepsilon_0\varepsilon_\perp\varepsilon_\parallel}H_0^2$ is the time-averaged Poynting vector that represents the energy flux direction and $H_0$ is the magnetic field strength of the transverse electromagnetic wave [60, 61]. At any incident direction $\theta_i$, negative refraction occurs in the hyperbolic frequency regime. One sees that negative refractive angle always occurs. In the MgS-$P6_3mc$ phase [Fig. 8(a)], at $\theta_i = 0.25\pi$, the refractive angle $\theta_r$ could drop to $-0.29\pi$ (at 5.4 THz). Similarly, for MgS-$P6_3/mmc$ [Fig. 8(b)], an incident angle of $0.25\pi$ would yield a refractive angle of $-0.18\pi$ at 7.5 THz. The situation of MgSe is plotted in Figs. 8(c) and 8(d). Such a refractive angle is on the same order as electron-contributed natural hyperbolic systems (near the visible light frequency regime) as predicted in previous works [62, 63]. These results indicate that clear and observable negative refractions can be observed in both anisotropic hexagonal and wurtzite phases, serving as potential THz waveguide and filter devices.

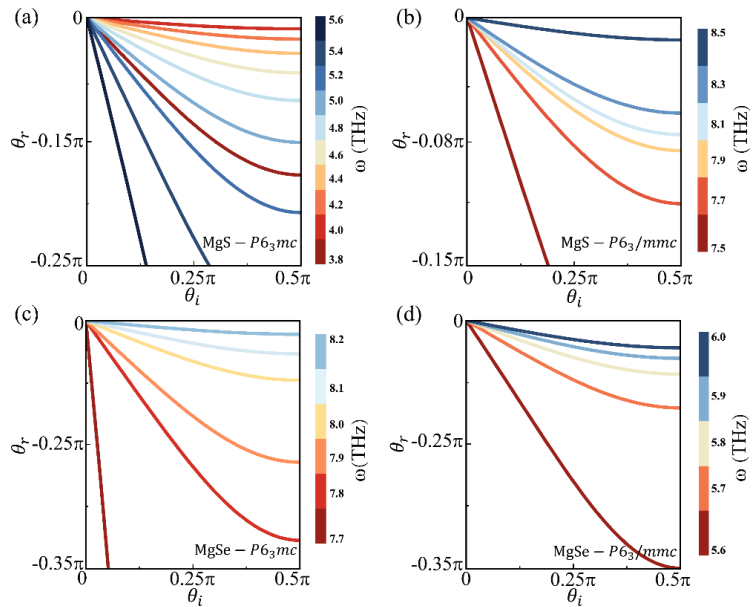

FIG. 8. Variation of negative refraction angle ($\theta_r$) with different incidence angles ($\theta_i$) of (a), (b) MgS and (c), (d) MgSe in $P6_3mc$ and $P6_3/mmc$ phases.

### IV. Conclusion

In summary, we use first-principles calculations that include anharmonic vibrational corrections to establish light and temperature dependent phase diagram of binary semiconductor MgS and MgSe compounds under nonresonant THz irradiation. We predict that the ground state $Fm\bar{3}m$ could exhibit nonvolatile phase transformations to $P6_3mc$ and $P6_3/mmc$ phases, owing to the larger dielectric function at low frequency. This indicates a noncontacting and non-destructive scheme for phase change, which could occur inside the bulk (not limited near the surface) of the system and the conventional nucleation-growth kinetics may be avoided. Furthermore, we propose that phononic natural hyperbolic dispersion exists in both anisotropic $P6_3mc$ and $P6_3/mmc$ phases, holding negative refraction of THz light without a propagation cutoff. The hyperbolic dispersion frequency window can be effectively modulated by temperature. Our mechanism and physical picture can be directly applied to similar phase transitions in other group IIA oxides, such as recently predicted BeO with significantly large ion-contributed dielectric constant. The phase transformation could also accompany with the change of polarization [64, 65], in a unconventional way. Considering the uncertainties (such as band gap underestimation and poor description of exciton binding energy) in predicting optical responses from electronic subsystems, the calculated phonon inspired optical feature can be directly compared with experimental observations. Our work presents a THz harnessed phase control to realize a high-performance exotic hyperbolic medium in simple and conventional semiconductors.

**Acknowledgments.** The work is supported by the National Natural Science Foundation of China (NSFC) under grant No. 12374065. The Hefei Advanced Computing Center is acknowledged where the calculations were performed.

**Data Availability.** The data that support the findings of this study are included in this article and are available from the corresponding author upon reasonable request.